# *J*-BAND SPECTROSCOPY OF THE *z* = 5.74 BAL QSO SDSSp J104433.04–012502.2[1]

R. W. GOODRICH[2], R. CAMPBELL[2], F.H. CHAFFEE[2], G. M. HILL[2], D. SPRAYBERRY[2], W. N. BRANDT[3], D. P. SCHNEIDER[3], S. KASPI[3], X. FAN[4], J. E. GUNN[5], AND M. A. STRAUSS[5]

Running Title: The *z* = 5.74 BAL QSO SDSS 1044–0125

## ABSTRACT

We use the NIRSPEC near-IR spectrometer on Keck II to obtain moderate resolution ($R = 1540$) spectroscopy that shows conclusively that the C IV emission line in the $z = 5.74$ quasar SDSSp J104433.04–012502.2 is accompanied by broad, blueshifted C IV absorption. The line has a "balnicity index" of 900 km s$^{-1}$ and a rest-frame equivalent width of 13.1 ± 1.3 Å relative to the continuum. This confirms its membership in the class of objects called "broad-absorption line" (BAL) QSOs. SDSSP J1044–0125 was previously suggested to be a BAL QSO based on its larger UV-to-X-ray flux ratio than most non-BAL QSOs. The C IV emission is of normal strength, implying a metallicity similar to that found in other, lower-redshift QSOs.

*Key words:* galaxies: active—galaxies: nuclei—quasars: absorption line—quasars: general—quasars: individual (SDSSp J104433.04–012502.2)

---

[1] Data presented here were obtained at The W. M. Keck Observatory, which is operated as a scientific partnership among the California Institute of Technology, the University of California, and the National Aeronautics and Space Administration. The Observatory was made possible by the generous financial support of the W. M. Keck Foundation.

[2] The W. M. Keck Observatory, 65-1120 Mamalahoa Highway, Kamuela, HI 96743.

[3] Department of Astronomy and Astrophysics, 525 Davey Laboratory, Pennsylvania State University, University Park, PA 16802.

[4] The Institute for Advanced Study, Princeton, NJ 08540.

[5] Princeton University Observatory, Princeton, NJ 08544.

## 1. INTRODUCTION

Recent developments in instrumentation and techniques have pushed forward our knowledge of the early Universe. New X-ray capabilities have extended our reach at high energies, while new surveys such as the Sloan Digital Sky Survey (SDSS; York et al. 2000) have produced a large number of high-redshift targets (e.g. Anderson et al. 2001).

At redshifts above 4 or 5 we can expect to learn much about the early evolution of our Universe, including how rapidly galaxies are formed and how quickly they are chemically enriched. Quasi-stellar objects (QSOs) provide bright beacons to this early phase of the Universe, and are powerful probes of local conditions at those times.

Fan et al. (2000) report the discovery that SDSSp J104433.04–012502.2 (hereafter SDSS 1044–0125) is a radio-quiet QSO at a reported redshift $z = 5.80$, showing emission lines such as N V and Si IV in data taken with the ESI spectrograph on Keck II. Follow-up XMM-Newton observations by Brandt et al. (2001) demonstrated that SDSS 1044–0125, while detected with high significance, is X-ray–weak relative to its optical flux when compared to most other QSOs. That paper suggested that SDSS 1044–0125 could be a broad-absorption line (BAL) QSO, which as a class have weaker observed X-ray fluxes (e.g. Green et al. 1995; Green & Mathur 1996; Gallagher et al. 1999), presumably caused by absorption within the BAL gas itself (Goodrich 1997). Brandt et al. further suggested that the best test of this hypothesis would be to observe C IV λ1550, redshifted to 1.05 μm in the observed frame, as this line shows strong absorption in nearly every BAL QSO.

Maiolino et al. (2001) recently obtained low-resolution infrared spectroscopy showing C IV absorption, and strengthening the case that SDSS 1044–1025 is a BAL QSO. However, their relatively low spectral resolution ($R \sim 75$, equivalent to 4000 km s$^{-1}$) did not allow them to come to a firm conclusion.

In the current *Letter* we present near-IR spectroscopy ($R = 1540$) of SDSS 1044–0125 that conclusively confirms the existence of strong C IV absorption in the near-IR. The C IV emission line strength, consistent with the emission lines observed in the far red by Fan et al. (2000), indicate that there has been significant chemical enrichment of at least the central regions of SDSS 1044–0125.

## 2. OBSERVATIONS

NIRSPEC (McLean et al. 1998) on the Keck II 10-m telescope was used on 2000 December 24 (UT) in its low-resolution mode. The detector is a 1024x1024 Aladdin-3 InSb array, with four amplifiers each reading out one quadrant of the detector. The wavelength region covered was from 9550–11,200 Å at a dispersion of roughly 2.15 Å/pixel and resolution of 3 pixels. This translates into a spectral resolution of $R = 1540$, or a velocity resolution of 195 km s$^{-1}$. The object was placed in one of five different positions along a 0.57 arcsec-wide slit 42 arcsec in length. A total of 19 exposures were taken, for a total integration time of 9660 sec. The night was photometric, and the data were flux-calibrated using SJ 9149. HR 4356 was used to correct for atmospheric absorption features.

The images were reduced using the NIRSPEC data reduction package *REDSPEC*. We corrected the images for bad pixels, spatially and spectrally rectified the images, and calculated and applied a wavelength scale. The latter was determined from the many night sky emission features (mostly due to OH) in the images. Note that these techniques are similar to those used in optical spectroscopy, with two exceptions: in many optical CCDs there is no great need to correct for bad pixels, and we used the usual infrared technique of using pair-subtracted images in order to better see and define the pixels containing the spectra. Long-slit sky subtraction was still performed to remove night sky residuals.

The spectra were then extracted in *VISTA* using an optimal extraction routine and a weighted mean was created. Note that one step missing was flat fielding. At the time of observation there were defects (dust and/or scratches) on the entrance window to the NIRSPEC dewar. The window is relatively near the telescope focal plane, and as a function of time, these defects rotated in and out of the slit. Hence flat fields taken at one time did not have the same slit illumination profile as the data. We decided that the data would be less degraded if they were *not* divided by a flat field. Instead a "pseudo-flat" was used that consisted simply of the relative gains of the two detector quadrants containing the data. This effectively removes a discontinuity where the spectrum crosses the boundary between the two quadrants.

The spectrum was scaled in flux by a factor of 0.69 to match the published Fan et al. (2000) spectrum between 9650 and 10,000 Å. The Fan et al. spectrum was originally scaled to match their $z'$ photometry.

## 3. RESULTS

Figure 1 shows the NIRSPEC spectrum, with the C IV emission line showing the classical combination of emission and absorption typical of BAL QSOs. The peak of the C IV emission was measured as $10,455 \pm 25$ Å, corresponding to a redshift of $z = 5.745 \pm 0.030$. This value is significantly lower than that reported by Fan et al. (2000; $z = 5.80 \pm 0.02$), mostly due to the higher signal-to-noise ratio in our spectrum (Fig. 1) and an ambiguity in Fan et al.'s identification of the N V and Ly$\alpha$ line peaks in the optical. (See the discussion in section 4.) Djorgovski et al. (2001) have concluded from high-quality optical spectra that $z = 5.73 \pm 0.01$, consistent with our value.[6]

The rest frame equivalent width (EW) of the entire C IV absorption (relative to the continuum only) is $13.1 \pm 1.3$ Å. This is similar to several of the BAL QSOs in the BQS (Bright Quasar Survey) studied by Brandt, Laor, & Wills (2000) and consistent with the value measured by Maiolino et al. (2001). Measured relative to the emission line (which is more consistent with the methodology of Brandt et al.) the rest EW is approximately 15 Å with a large uncertainty due to the unknown underlying emission line shape.

The C IV emission feature is partially absorbed by the C IV absorption, so we base the following on measurements of just the line profile redward of the peak and assuming a symmetric line profile. The rest EW of emission is $19.9 \pm 2.0$ Å and the FWHM is 6000 km s$^{-1}$. These values are typical for radio-quiet QSOs, and in particular for high-redshift QSOs (Fan et al. 2001). The relatively unabsorbed emission profile just blueward of the peak indicates that the emission line likely is not symmetric, so the calculations above may slightly underestimate the true emission EW.

Weymann et al. (1991) invented the term "balnicity" to attempt to quantify the strength of absorption of the BAL lines. The balnicity of the C IV absorption feature in SDSS 1044–0125 is only 900 km s$^{-1}$. This is smaller than might seem reasonable from Fig. 1, given the equivalent widths of absorption vs. emission, until one recalls that the definition of balnicity (see Weymann et al. 1991 for the full definition) ignores absorption within the first 5,000 km s$^{-1}$ blue of the peak (note the black bar at the bottom of Fig. 1, representing the range over which balnicity is calculated in SDSS 1044–0125). In SDSS 1044–0125 most of the absorption occurs within this first 5,000 km s$^{-1}$.

In Figure 1, at 10,500 Å on the red side of the C IV emission line are two narrow absorption features. Their wavelengths are measured to be 10,499 and 10,516 Å, a velocity splitting of $504 \pm 50$ km s$^{-1}$ and separation from the peak of 1500 km s$^{-1}$. This is almost surely a C IV doublet, which has an intrinsic velocity splitting of 498 km s$^{-1}$ and a line ratio similar to that observed. Redshifted absorption is not unheard of in BAL QSOs (e.g. Glenn, Schmidt, & Foltz 1994). Another, blueshifted C IV doublet appears at $\lambda\lambda$10,035, 10,051 ($493 \pm 50$ km s$^{-1}$ splitting), separated by –12,000 km s$^{-1}$ from the C IV emission peak.

## 4. DISCUSSION

C IV absorption is normally associated with absorption in other high-ionization UV resonance lines, such as Si IV, N V, and O VI. Indeed, in the original Fan et al. ESI spectrum (see also Fig. 1) absorption can be seen blueward of Si IV. Only with the perspective of the C IV absorption can we be sure that this is likely a Si IV BAL and not some undercorrected atmospheric absorption feature or series of intergalactic absorption features. N V can be problematic because of the adjacent, and much stronger Ly$\alpha$ emission line. Often the N

---

[6] Unfortunately, Iwata et al. (2001) used the incorrect, higher redshift of 5.80 to search, unsuccessfully, for CO emission. Their bandpass did not include the predicted frequency for CO emission at $z = 5.73$ or 5.745.

V absorption absorbs much of the Lyα emission, and this certainly could be the case in SDSS 1044–0125. In fact the emission peak seen at 8350 Å by Fan et al. matches the expected wavelength of N V. A much smaller emission peak around 8200 Å matches better the wavelength of Lyα (Fig. 1), and indicates that much of the Lyα emission is indeed absorbed by the N V BAL trough. The O VI line is weaker and cut into by numerous intervening Lyα absorption systems. Djorgovski et al. (2001) have arrived at a similar conclusion from high-resolution and high signal-to-noise ESI spectra in the optical.

Brandt et al. (2001) pointed out that the X-ray weakness of SDSS 1044–0125 could arise if it is a BAL QSO. They further suggested that spectroscopy of the C IV region at 10,500 Å would resolve the issue. Our NIRSPEC spectra confirm that the object is indeed a BAL QSO, and one with emission and absorption parameters that are normal for lower redshift objects of the BAL class.

Note that due to the high redshift of SDSS 1044–1025, the X-ray observations access relatively high energies (up to ~50 keV in the rest frame). Hence the constraint on the amount of absorption needed to explain the paucity of X-rays is severe. Brandt et al. estimate that a column density of at least $10^{24}$ cm$^{-2}$ is required, and that partial covering is needed. If the X-ray and UV absorbers are closely related, this would imply highly saturated UV line absorption, but due to the partial covering of the source the flux in the absorption would not necessarily go to zero (e.g. Goodrich 1997).

The fact that the highest redshift QSO known is a BAL QSO hints that such QSOs could be a larger fraction of the total at high redshift than at lower redshifts. There are two reasons why we might expect to find a larger fraction of BAL QSOs at high redshift. First, BAL QSOs may represent a generally younger stage in the evolution of a QSO, one in which the surrounding cocoon of gas and dust has not yet been thoroughly cleared. At low redshift Mrk 231 is likely an example of this stage (e.g. Goodrich & Miller 1994).

Second, Goodrich (1997) has argued for a qualitative difference in the absorption characteristics of AGN at high vs. low luminosity. Essentially the dramatic deep, high-velocity absorption seen in BAL QSOs is not seen in the low-luminosity counterparts, Seyfert 1s. If BAL absorption increases with increasing luminosity, we would expect a flux-limited high-redshift sample of QSOs to show more BAL QSOs than a corresponding low-redshift sample (e.g. Goodrich 1997). Alternatively, there may exist a luminosity above which the BAL-like outflows "turn on;" above this luminosity the fraction of BAL QSOs versus luminosity could remain constant.

Anderson et al. (2001) analyzed QSOs in the redshift range from 4 to 5 and found no difference between the fraction of BAL QSOs compared to lower redshift surveys. The discovery of more even higher-redshift QSOs will allow us to extend this test in the future.

The simple fact that we see emission line equivalent widths and a continuum shape (Maiolino et al. 2001) that are similar to low-$z$ QSOs indicates that the emission line gas in SDSS 1044–0125 has a similar metallicity (e.g. Hamann & Ferland 1999), presumably solar or higher. Discussed by Fan et al. (2000), this is a remarkable fact that is worth repeating. Even at such redshifts approaching $z = 6$ we are seeing chemically enriched gas in the nuclei of (presumably) galaxies! This indicates that at least in some galaxies chemical enrichment occurs on a very short time scale. A selection effect may be at work here as well. If only a handful of galaxies are rapidly enriched so early in the life of the Universe, but this enrichment is a prerequisite to the formation of a supermassive black hole (or is somehow associated with the hole formation), then by observing QSOs we would only select those objects in which the enrichment has already occurred.

Powerful surveys such as the SDSS and the second Palomar Sky Survey (Kennefick, Djorgovski, & de Carvalho 1995) are discovering dozens of $z > 4$ QSOs each year. We need to push our spectroscopy ever further into the infrared in order to study the UV resonance lines that provide so much useful diagnostic information. For example, a comparison of the frequency of BAL QSOs with luminosity may require C IV spectroscopy; BAL features are generally weaker in Si IV, and more ambiguous in N V. For $z > 5$, C IV is most readily observed with near-IR spectrographs, something which has until recently not been attempted.

The authors wish to extend special thanks to those of Hawaiian ancestry on whose sacred mountain we are privileged to be guests. Without their generous hospitality, none of the observations presented herein would have been possible. WNB thanks the Alfred P. Sloan Foundation for financial support of this work. DPS acknowledges the support of NASA grant AST99-00703, MAS acknowledges the support of NSF grant AST-0071091, and XF acknowledges NSF grant PHY00-70928.

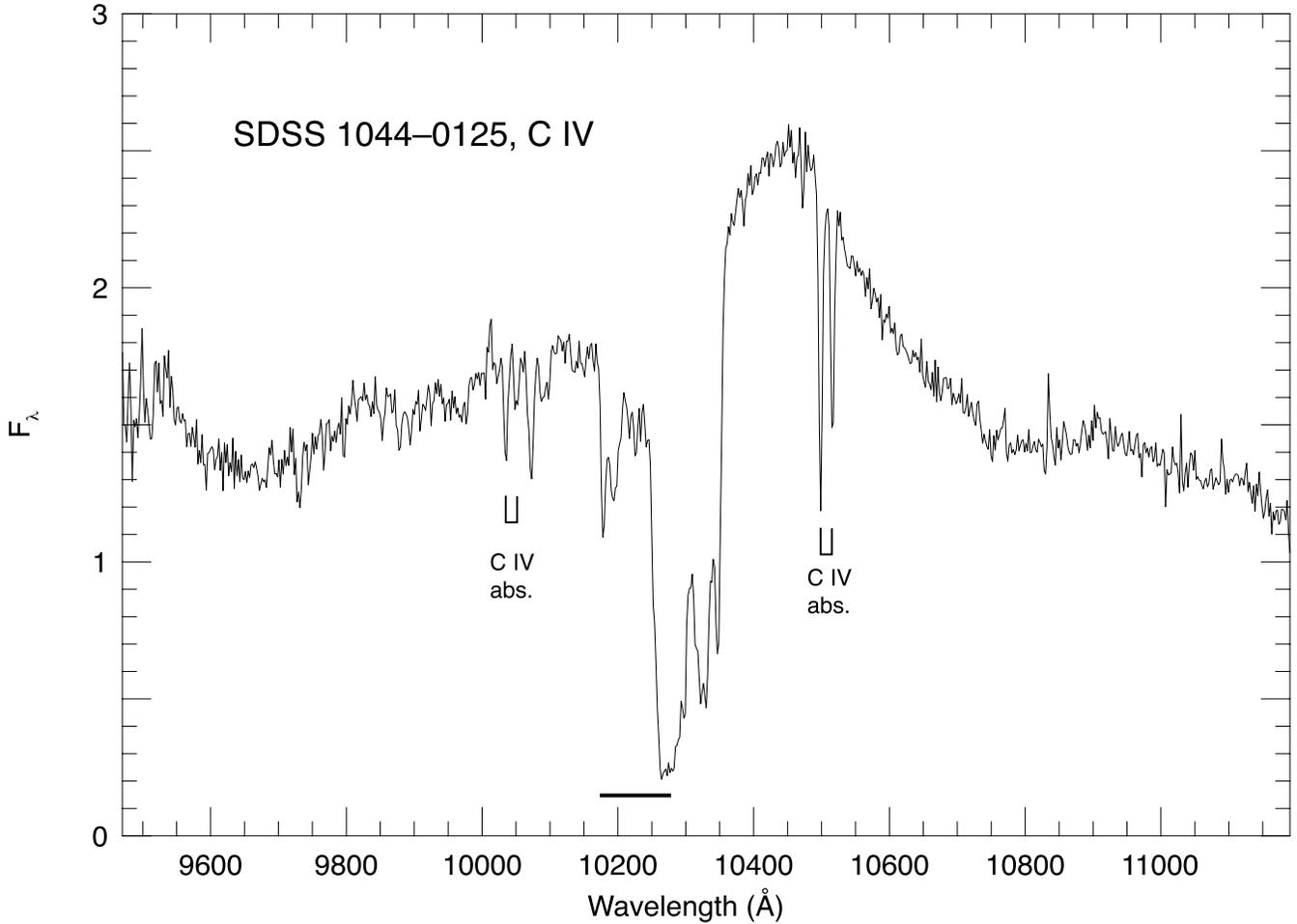

Figure 1 — The C IV emission and absorption features as seen in near-IR spectroscopy taken with NIRSPEC. The flux density is in units of $10^{-17}$ erg s$^{-1}$ cm$^{-2}$ Å$^{-1}$, and the wavelength scale is in Å. Note the broad, blueshifted absorption characteristic of a broad-absorption line (BAL) QSO. The two deep absorption features to the red of the emission peak (near 10,500 Å) are also a C IV doublet. The spectral resolution of this spectrum is $R = 1540$, corresponding to a velocity resolution of 195 km s$^{-1}$. The dark bar below the C IV absorption trough represents the velocity range $-8100$ to $-5000$ km s$^{-1}$, over which the "balnicity index" is calculated. The definition of "balnicity" precludes measurement of the deep absorption at smaller velocities.

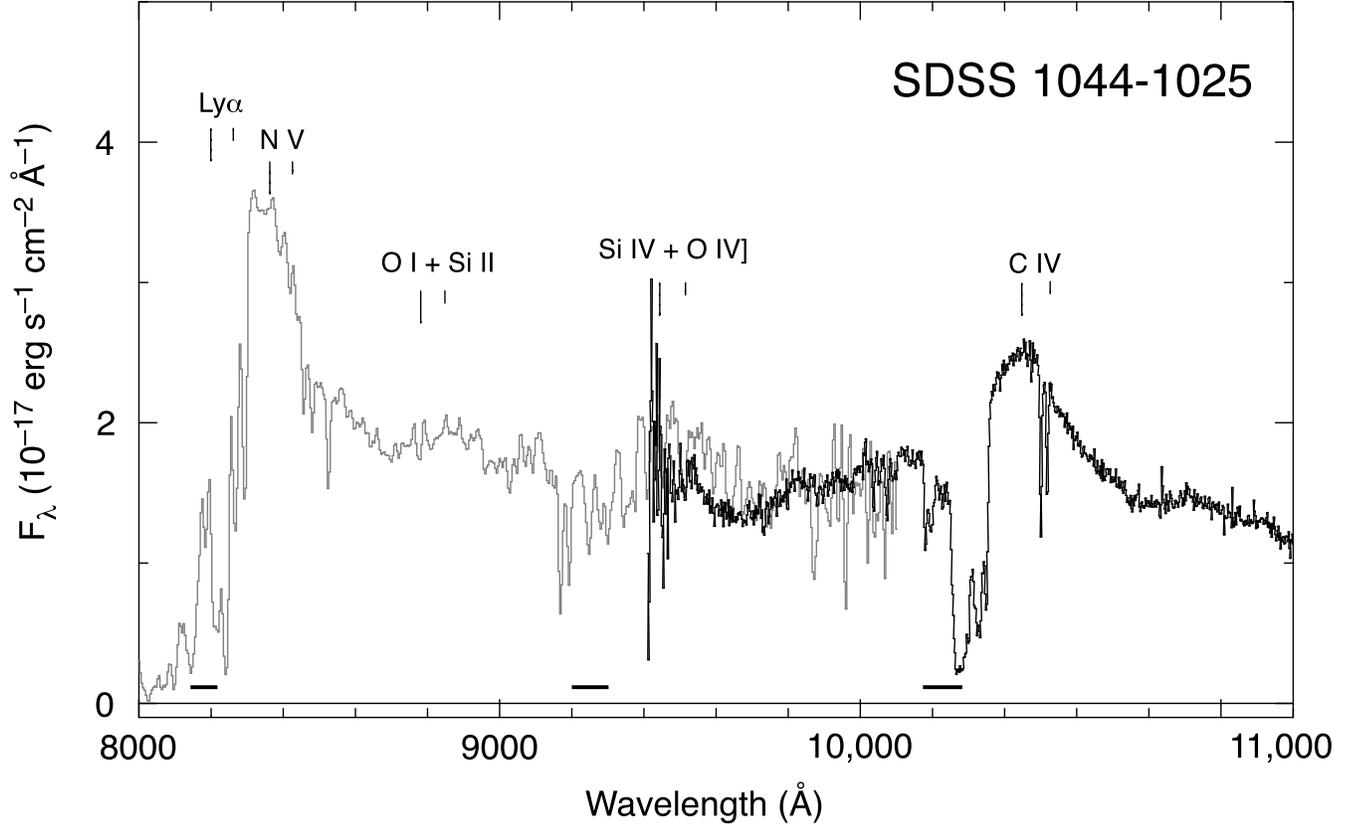

Figure 2 — The composite ESI and NIRSPEC spectrum. Expected emission line wavelengths are shown for $z = 5.74$ (long tick marks) and $z = 5.80$ (short tick marks). Note that the mismatch near Si IV between the two spectra is probably due to uncertainties in calibrating the edges of the NIRSPEC spectra. The signal-to-noise is much higher in the NIRSPEC spectrum than in the ESI spectrum, so while NIRSPEC has detected some real absorption features, in this region ESI is dominated by noise. As in Fig. 1, the horizontal bars at the bottom of the plot represent the velocity range $-8100$ to $-5000$ km s$^{-1}$ for (left to right) N V, Si IV, and C IV absorption features.